\documentclass[fleqn,twoside]{article}
\usepackage{espcrc2}
\usepackage{epsfig}

\usepackage{amssymb}
\usepackage{graphicx}
\usepackage{dcolumn}
\usepackage{bm}

\newcommand{\ttbs}{\char'134}
\newcommand{\AmS}{{\protect\the\textfont2
  A\kern-.1667em\lower.5ex\hbox{M}\kern-.125emS}}

\title{Force between a magnetic tip  and an inhomogeneous superconductor at zero  field 
\tt\ttbs\thanks {Research supported in part by CNPq-Brazil.}}

\author{Gilson Carneiro
\address{Instituto de F\'{\i}sica, Universidade Federal do Rio de Janeiro,  
C.P. 68528, 21941-972, Rio de Janeiro-RJ, Brasil }}

\begin{document}

\begin{abstract}

{The force exerted by a semi-infinite inhomogeneous superconductor  with a planar interface to vacuum on a magnetic tip  is studied  theoretically in the absence of external magnetic fields. It is shown that the force has a contribution from inhomogeneities due to material defects with  unique characteristics. Defects are taken into account in the London limit by allowing the  mass parameter  to vary spatially. The contribution from defects to the force is calculated analytically to first order in the deviation of the mass parameter from its constant value for the homogeneous superconductor, assuming that the tip is a point dipole perpendicular to the interface, and that it does not spontaneously create vortex matter. Random point defects and linear localized defects are considered phenomenologically. For each defect type the force dependence on the dipole position coordinates is obtained, and the force magnitudes are estimated numerically. The predictions for the dependence of the  linear  defect force on the dipole lateral position are found to agree  qualitatively with experiment.}

\vspace{1pc}
\end{abstract}


\maketitle


A magnet and a superconductor placed in the vicinity of each other interact through a force that results  from the changes in the superconducting state induced by the magnet. Since inhomogeneities in the superconductor due to material defects modify the magnet-superconductor interaction, the force must contain contributions from  defects. This contribution must be  present in the force on a magnetic tip  measured  by the technique of magnetic force microscopy (MFM) \cite{mfm1} applied at zero external field, and can, in principle, be measured if its  magnitude  is larger than the MFM resolution, about  $ 1\, pN$.
 
For planar superconductor-vacuum interfaces, as assumed here, translational symmetry considerations alone suggest that there is a  contribution from defects  to the force on the tip with unique characteristics.  For defect-free materials, the tip-superconductor interaction energy is invariant under translations parallel to the interface, and depends only on the tip-superconductor distance. This  implies that  the force on the tip, equal to the gradient of the  interaction energy, is perpendicular to the interface and independent of the tip lateral position coordinates. This is no longer true if the superconductor has defects, because translational invariance is broken. The tip-superconductor interaction energy now depends also on the tip lateral position, and  so does the force, which is no longer perpendicular to the interface. Thus, the force on the tip  has a contribution from defects with components parallel and perpendicular to the interface, both dependent on the tip lateral position as well as on the tip-superconductor distance. This contribution  can, in principle, be singled out in MFM measurements since it is responsible for the lateral force component and for the dependence of both  force components on the tip  lateral tip position. The objective of this  paper is to study the properties of  the contribution from defects to the force and to estimate its magnitude using a simple  model. 

The general framework for the description of inhomogeneous superconductors is the Ginzburg-Landau (G-L) model with  spatially varying coefficients of the G-L free-energy  expansion \cite{rev1}. Here the G-L model is used in the London limit, in which case only the mass parameter of the  G-L expansion appears. The effect of the tip magnetic field on the superconductor is to induce a screening current  and  create vortex matter, when the tip is close enough to the interface.
Here the defect contribution to the force is calculated assuming that the tip does not create vortex matter. This is the situation of greatest interest because, as shown here, the defect contribution results solely from to the spatial dependence of the mass parameter. 

\begin{figure}[t]
\centerline{\includegraphics[scale=0.2]{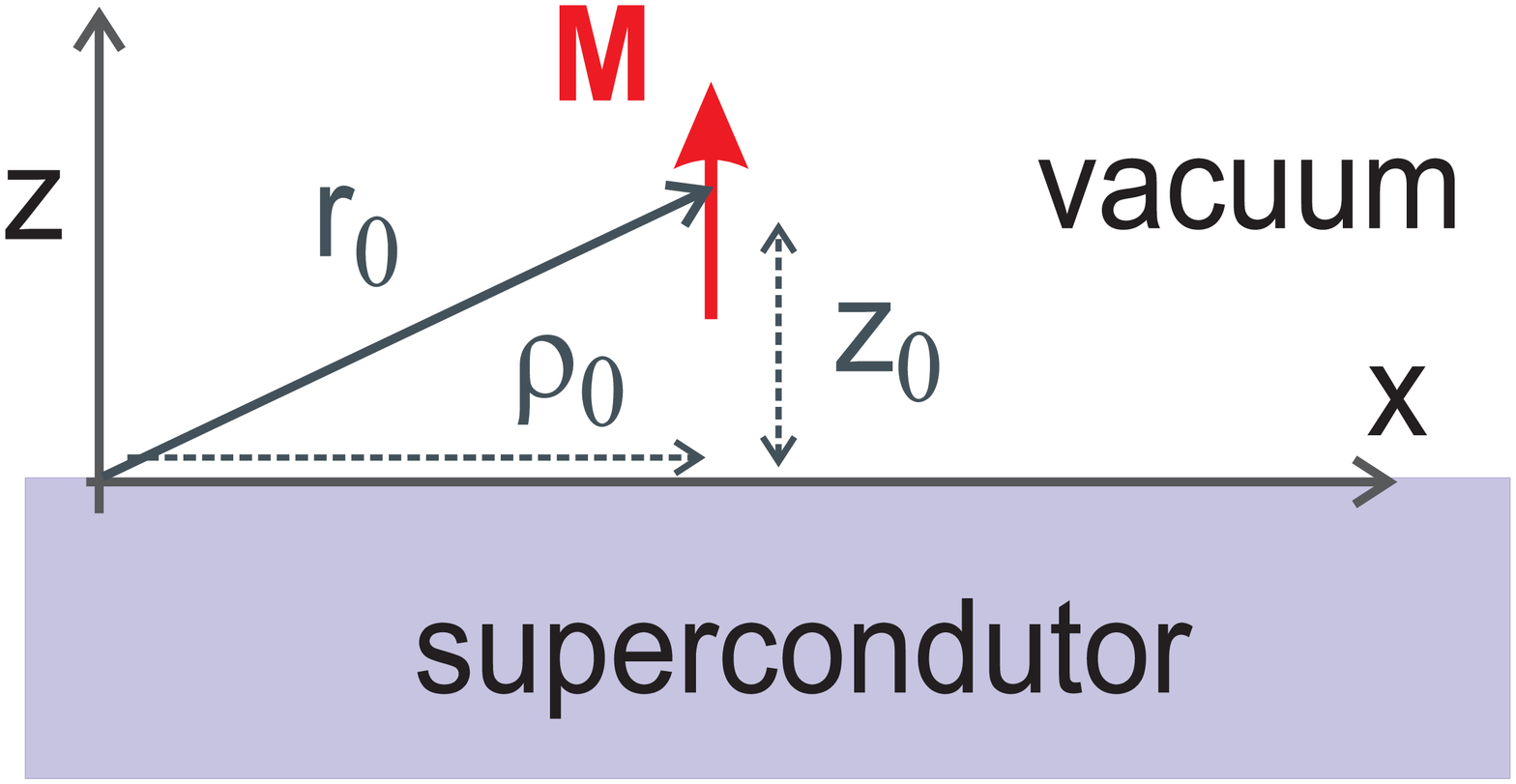}}
\caption{Schematic view of semi-infinite superconductor and  point dipole ${\bf M}=M_z\hat{{\bf z}}$, at  ${\bf r}_0 = ({\mbox{\boldmath $\rho$}}_0,z_0)$.}
\label{fig.fig1}
\end{figure}

The model considers a semi-infinite isotropic superconductor, separated from the vacuum by a planar interface parallel to the $x-y$ plane, occupying  the region $-\infty\leq z \leq 0$.  The  magnetic tip is approximated by a point dipole perpendicular to the interface, ${\bf M}=M_z\hat{{\bf z}}$,  placed  at ${\bf r}_0 =(x_0,y_0,z_0)\equiv ({\mbox{\boldmath $\rho$}}_0,z_0>0)$
(Fig.\ \ref{fig.fig1}).
The free-energy  in the London limit, assuming that the superconductor is vortex matter free, is given by \cite{rev1}
\begin{equation} 
G = \int d^2\rho\int_{-\infty}^0 dz \frac{\mid{\bf A}\mid^2}{8\pi\lambda^2}
+ \int d^3r \frac{\mid {\mbox{\boldmath ${\nabla}$}}\times {\bf A}\mid^2}{8\pi}\;.
  \label{eq.fel}
  \end{equation}
where ${\bf r}\equiv  ({\mbox{\boldmath $\rho$}},z)$,  ${\bf A}({\bf r})$ is the vector potential, and $\lambda$ is the penetration depth, related to the G-L mass  parameter, $m$,  by $\lambda^2=mc^2/(4\pi n_se^{*2})$, where $e^*=2e$, and $n_s$ is the density of Cooper pairs. For homogeneous superconductors the mass parameter is a constant, $m_0$, and so is the penetration depth, $\lambda_0$. Inhomogeneities are described by   $m$ dependent on the spatial coordinates,   $m({\mbox{\boldmath $\rho$}},z)=m_0 + \delta m({\mbox{\boldmath $\rho$}},z)$.
Here it is assumed that  $\delta m/m_0 \ll 1$. The free-energy  to  first order in $\delta m/m_0$ is written as
\linebreak $G^{(1)}({\bf r}_0) = G^{(0)}(z_0) + \delta G({\bf r}_0)$. The first term, $ G^{(0)}$, is the homogeneous superconductor free-energy, given by Eq.\ (\ref{eq.fel}) with $m=m_0$,  and ${\bf A}$  equal to  the equilibrium vector potential in the absence of defects, denoted ${\bf A}_0({\bf r})$. It  depends  only on $z_0$ because of translational invariance parallel to the interface. The second term is the first order correction, given by
 \begin{equation} 
\delta G({\bf r}_0) = -\int d^2\rho\int_{-\infty}^0 dz \frac{\mid{\bf A}_0({\mbox{\boldmath $\rho$}},z)\mid^2}{8\pi\lambda^2_0}
\frac{\delta m({\mbox{\boldmath $\rho$}},z)}{m_0}
\;.
  \label{eq.dfe}
  \end{equation}
The correction to first order in $\delta m/m_0$ to the free energy, Eq.\ (\ref{eq.dfe}), is just the change in the kinetic energy of the homogeneous superconductor  caused by the inhomogeneity, since the screening current is given by  ${\bf j}_{sc}=-(n_se^{*2}/m_0c){\bf A}_0$.   The force on the dipole is  ${\bf F}^{(1)}({\bf r}_0)  =  F^{(0)}(z_0)\hat{\bf z}+\delta{\bf F}({\bf r}_0)$ where the first term is the homogeneous superconductor contribution with $F^{(0)}(z_0)=-dG^{(0)}(z_0)/dz_0 $ and 
$\delta{\bf F}({\bf r}_0)=-{\mbox{\boldmath ${\nabla}$}}_0 \delta G({\bf r}_0)$ is the contribution from defects. 
 
\begin{figure}[t]
\centerline{\includegraphics[scale=0.25]{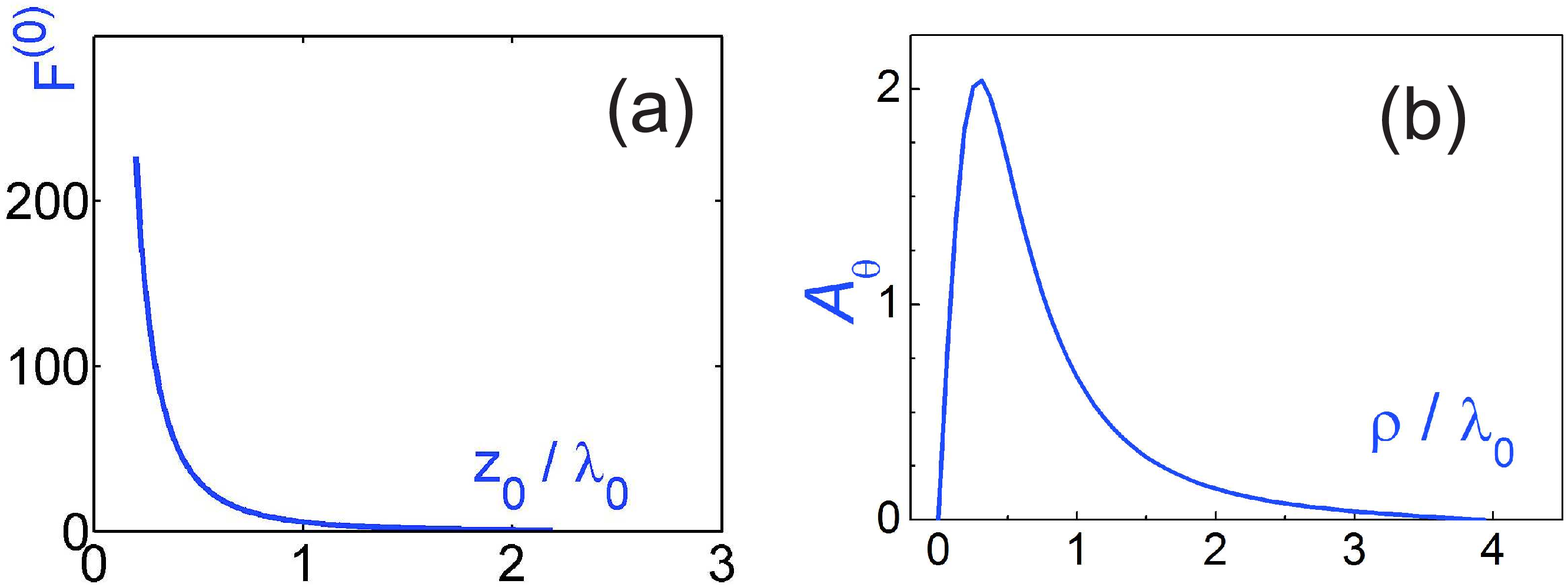}}
\caption{a) Dependence of $F^{(0)}(z_0)$ on $z_0$ ($F^{(0)}$ in units of $\epsilon_0=(\phi_0/4\pi\lambda_0)^2$) for $M_z/\phi_0\lambda_0=1.0$. b) Dependence of $A_{\theta}(\rho,z;z_0)$ on $\rho$, for $\rho_0=0,\, z=-\lambda_0/8$,  $z_0=0.3 \lambda_0$, and $M_z / \phi_0\lambda_0=1.0$ ($A_{\theta}$ in units of $\phi_0/\lambda_0$). }
\label{fig.fig2}
\end{figure}

The homogeneous superconductor free-energy is  $G^{(0)}(z_0)=-M_z\,b'_z(z_0)/2$,  where $b'_z$ is the $z$-component of the magnetic field of the dipole screening supercurrent at the position of the dipole \cite{gmc1}. The calculations carried out here use the analytic expressions for $ b'_z$, and  ${\bf A}_0$ obtained in Ref.\cite{gmc1}. The force  $F^{(0)}(z_0)$ is repulsive ($F^{(0)}(z_0)>0$) and increases as $z_0$ decreases, typically as shown in Fig.\ \ref{fig.fig2}. For the dipole at $({\mbox{\boldmath $\rho$}}_0=0,z_0>0)$, ${\bf A}_0$ is given by ${\bf A}_0({\mbox{\boldmath $\rho$}},z;z_0)=A_{\theta}(\rho,z;z_0)\,\hat{{\mbox{\boldmath ${\theta}$}}}$, and $\hat{{\mbox{\boldmath ${\theta}$}}}$ is the unit vector in the direction of the polar angle. For the dipole located  at  ${\mbox{\boldmath $\rho$}}_0 \neq 0$,  the vector potential is ${\bf A}_0({\mbox{\boldmath $\rho$}} - {\mbox{\boldmath $\rho$}}_0, z; z_0)$, because of  translational invariance parallel to the interface. A typical  dependence of  $A_{\theta}$ on $\rho$ at constant $z$ is shown in Fig.\ \ref{fig.fig2}. For other values of $z_0$ and $z$,  $A_{\theta}$ decreases smoothly as $\mid z \mid$ increases, vanishing for $\mid z \mid \gg \lambda_0$, and also decreases with increasing $z_0$. 

Next, the contribution from defects to the force is calculated for  random point defects and  linear localized defects, assuming in each case that the spatial dependence of $\delta m({\mbox{\boldmath $\rho$}},z)$ is  know.   

{\it Random Point Defects.} In this case $\delta m({\bf r})$ is a  random function with zero average satisfying\linebreak 
$<\delta m({\bf r})\delta m({\bf r}')>  =  \gamma_m\,\delta({\bf r}-{\bf r}')$, where $< >$ denotes average over the randomness \cite{rev1}. According to Eq.\ (\ref{eq.dfe}), the force is a random function of the dipole lateral position, ${\mbox{\boldmath $\rho$}}_0$, with zero average and root mean square (RMS) fluctuation given by  
\begin{equation} 
<\mid \delta  F_{\alpha}\mid ^2>^{1/2} =   2\pi\epsilon_0 (\frac{\gamma_m}{m^2_0\lambda^3_0})^{1/2}
(\frac{M_z}{\phi_0\lambda_0})^2 R_{\alpha}(z_0)  \; ,
  \label{eq.fr1}
  \end{equation}
where $\alpha$ labels the force components (lateral: $\alpha=\rho $, perpendicular: $\alpha=z$),  
$ \epsilon_0=(\phi_0/4\pi\lambda_0)^2$,  and $R_{\alpha}$ are dimensionless functions, independent of $M_z$,  given by:

\noindent $ R_{\rho}(z_0)  = \lambda^{-2}_0 \int d^2\rho\,\int^{0}_{-\infty} dz
\mid {\mbox{\boldmath ${\nabla}$}}_{\rho} a^2_0({\mbox{\boldmath $\rho$}}, z; z_0)\mid^2 $;

\begin{figure}[t]
\centerline{\includegraphics[scale=0.3]{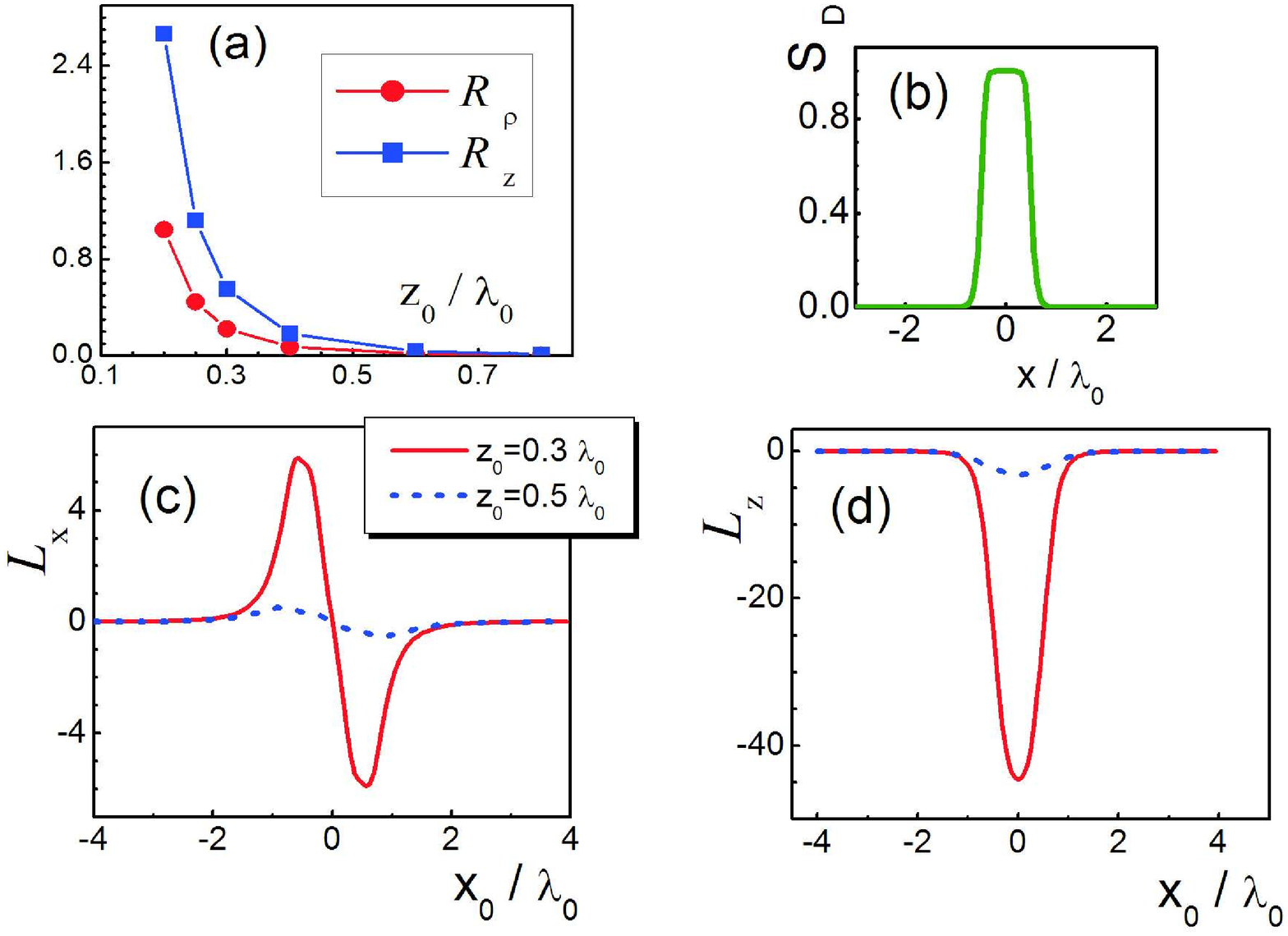}}
\caption{(a) Random defects functions $R_{\rho}(z_0)$ and $R_{z}(z_0)$ vs. $z_0$. (b) Linear defect shape function for $a_D=0.5\lambda_0$, and $n=5$ vs. dipole lateral position $x_0$. Linear defect functions $L_{x}$ (c), and $L_z$ (d) vs. $x_0$ and $z_0$.}
\label{fig.fig3}
\end{figure}

\noindent $ R_{z}(z_0)  = \lambda^{-2}_0 \int d^2\rho\int^{0}_{-\infty} dz
\mid  \partial a^2_0({\mbox{\boldmath $\rho$}}, z; z_0)/\partial z_0\mid^2$, 

\noindent where  ${\bf a}_0$, defined by  ${\bf A}_0 \equiv (\phi_0/\lambda_0) (M_z/\phi_0 \lambda_0) 
{\bf a}_0$, is dimensionless.  The functions $R_{\rho}(z_0)$ and $R_{z}(z_0)$ decrease smoothly with increasing $z_0$, as shown in Fig.\ \ref{fig.fig3}a.  For a  numerical estimate of $<\mid \delta  F_{\alpha}\mid ^2>^{1/2}$ the value of $\gamma_m$ is taken from  Ref.\cite{rev1}, obtained from collective vortex pinning theory  at low temperatures, namely $\gamma_m/(m^2_0\xi^3)\sim 10^{-2}$, where $\xi$ is the vortex core radius. Thus $\gamma_m/(m^2_0\lambda^3_0)\sim 10^{-2}(\xi/ \lambda_0)^3$. Assuming $\lambda_0 = \,150\,nm$, $\lambda_0/\xi=10$ and $z_0\sim 0.3 \lambda_0$, for which  $R_{\rho},\;R_z \sim 1$ (Fig.\ \ref{fig.fig3}a), gives  $(< (\delta F)^2>)^{1/2}\sim 2\times 10^{-2}\epsilon_0 (M_z/\phi_0 \lambda_0)^2$ \linebreak $\sim 2\times 10^{-1}(M_z/\phi_0 \lambda_0)^2\; pN$, for both the lateral and perpendicular force components. Thus, the predicted value of $(< (\delta F)^2>)^{1/2}$ are larger than $ 1\, pN$ for $M_z/\phi_0 \lambda_0 >2$.

{\it Localized Linear Defect.} In this case $\delta m({\bf r})$ is written  as $\delta m({\bf r})= m_0 \chi_m S_D({x})$, where\linebreak $S_D(x)=[1+(x/a_D)^{2n}]^{ -1}$ is the defect shape function shown in Fig.\ \ref{fig.fig3}b, $ \chi_m $ a dimensionless constant, $a_D$ the defect length scale,  and $n$ is a positive integer, chosen large  in order  that  the boundary between the defect and the homogeneous superconductor is sharp. The assumption that $\delta m/m_0 \ll 1$ is equivalent $ \mid \chi_m \mid \ll 1 $. 
The force is given by
\begin{equation} 
\delta {\bf F}(x_0, z_0) =  2\pi \epsilon_0
(\frac{M_z}{\phi_0\lambda_0})^2 \chi_m{\bf L}(x_0, z_0)  \; ,
\label{eq.fd1}
\end{equation}
where   ${\bf L}=(L_x,L_z)$ is a  dimensionless vector  defined as: 
 
\noindent  ${\bf L}=\lambda^{-2}_0{\mbox{\boldmath ${\nabla}$}}_0 \int d^2\rho \int^{0}_{-\infty} dz  
 a^2_0({\mbox{\boldmath $\rho$}}, z; z_0)  S_D(x+x_0,z)$.
 
\noindent Typical results for $L_{x}$ and $L_z$ as functions of the dipole position coordinates, $x_0$ and $z_0$ are shown in Fig.\ \ref{fig.fig2}c and Fig.\ \ref{fig.fig2}d. 
The highlights of these figures are the following. The functions  $L_{x}$ and $L_z$ are non-vanishing only in a range of  $\mid x_0 \mid $ values   comparable $a_D$, and decrease fast with increasing $z_0$. The former depends on the value chosen for $a_D$, and is true in general, except for  $a_D\ll \lambda_0$. In this case the ranges of $L_{x}$ and $L_z$ are comparable to $\lambda_0$, because the range of ${\bf A}_0$ is $\lambda_0$, but their magnitudes  are considerably reduced. The function $L_{x}$ oscillates between positive and negative values, whereas $L_z$ is always negative. This behavior results  because $L_x$ is related to the derivative of $S_D$ and $L_z$ to  $S_D$ itself, and from the functional dependence of $S_D$ chosen here. 
For an order of magnitude estimate of the force, the same parameters used in the case of random point defects are adopted, and $\chi_m \sim 10^{-1}$ is chosen. The result is 
$ \delta F_{\alpha}\sim 2\pi (M_z/\phi_0 \lambda_0)^2\,L_{\alpha}\; pN$, where $ \alpha=x,\; z$. The maximum values of the force components for  $z_0=0.3\lambda_0$, using the results shown in Fig.\ \ref{fig.fig3},  are $ \delta F_{x} \sim 40(M_z/\phi_0 \lambda_0)^2\, pN$, and $\delta F_{z} \sim 3\times 10^2(M_z/\phi_0 \lambda_0)^2\, pN$. Thus, for  $z_0=0.3\lambda_0$,  force values $> 1\, pN$ are predicted for modest values of $M_z/\phi_0 \lambda_0$.

The predictions for the force from linear defects is now compared to MFM force measurements at zero field reported in Ref.\cite{mfm2}. These  measurements find that the perpendicular component of the force in the vicinity of a stacking fault dislocation in the superconducting material depends on the tip lateral position (see Fig.3 and Fig.4 of Ref.\cite{mfm2}) in a manner similar to that predicted here for a linear defect 
(Fig.\ \ref{fig.fig3}d). The interpretation of these measurements  based on the results of this paper is that the  force behavior is due to the spatial dependence of mass parameter caused by the stacking fault dislocation. 
The latter statement is justified by the work reported in Ref.\cite{kml} showing that defects in the crystal lattice of the superconducting material cause  spatial dependencies of the G-L parameters, and by the fact that the stacking fault dislocation in the sample used in Ref.\cite{kml} has an approximate linear symmetry.  The authors of  Ref.\cite{mfm2} interpret their results  differently.  They attribute the observed behavior to vortices created by unshielded  magnetic fields, from the earth and from the experimental apparatus, pinned by the stacking fault dislocation.  Both interpretations  account qualitatively for the spatial dependence of the force. One possible way to distinguish between them experimentally would be to revert  the tip magnetic moment and observe how the force changes. The contribution from defects to the force remains unchanged, whereas that from vortices changes sign. 

In conclusion then this paper  shows that defects in the superconducting material, as described by spatially dependent G-L mass parameters, give contributions  to the force on the dipole with  dependencies on the dipole   lateral position characteristic of the type of defect. The order of magnitude  estimates carried out here indicate that the force from random point defects is probably too small to be observed experimentally. On the other hand, the force from a linear localized defect is estimated to be larger  than the resolution of the  MFM technique, and may already been observed. 

The calculations carried out here neglect creation of vortex matter by the dipole.
This  happens when the tip is  close enough to the interface, because the dipole magnetic field inside the superconductor becomes  large enough to create  vortex half loops \cite{gmc1}. The presence of vortex matter changes the defect contribution to the force on the dipole. It is expected that, to first order in 
$\delta m/m_0$, it reduces the magnitude of this contribution. The reason is that the defect force results from the change in the kinetic energy caused by the inhomogeneity, and the vortex matter created by the dipole generates a current that opposes the dipole screening current, reducing the total current and, consequently, the kinetic energy.   The question is how large this reduction is.  To answer it requires considerations beyond the scope of this paper. Work along these lines in under way and will be reported elsewhere.


\begin{thebibliography}{9}

\bibitem{mfm1} A. Moser, H. J. Hug, I. Parashikov, B. Stiefel, O. Fritz, H. Thomas, A.
Baratoff,  H.-J. G\"{u}ntherodt, and P. Chaudhari, Phys. Rev. Lett. {\bf 74}, 1847  (1995).

\bibitem{rev1} For a review see: G. Blatter, M. V. Feigel\'{}man, V. B. Geshkenbein, A. I.
Larkin, and V. M. Vinokur, Rev. Mod. Physics {\bf 66}, 1125 (1994), and references therein.

\bibitem{gmc1} G. Carneiro, Phys. Rev. B {\bf 69}, 214504 (2004).

\bibitem{mfm2} U. H. Pi,  Z. G. Khim, D. H. Kim, A. Schwarz, M. Liebmann, and R. Wiesendanger, Phys. Rev. B {\bf 69}, 094518 (2004).

\bibitem{kml} E. Schneider,and H. Kronm\"{u}ller, Phys. Stat. Sol. {\bf 74}, 261 (1976), and references therein.

\end{thebibliography}
\end{document}